\documentclass[11pt,a4pape,rpdftex]{article}
\usepackage{amsmath, amscd, amssymb, amsthm, latexsym
} \usepackage{float}

\usepackage[colorlinks=true,linkcolor=red]{hyperref}
\usepackage{hyperref} 
\usepackage{tikz,pgfplots}
\usetikzlibrary{positioning}

\newtheorem{theorem}{Theorem}
\newtheorem{proposition}[theorem]{Proposition}
\newtheorem{lemma}[theorem]{Lemma}
\newtheorem{example}[theorem]{Example}

\newtheorem{corollary}[theorem]{Corollary}

\newcommand{\means}[1]{\hBox{$ [\kern -.4em [\, {#1}\, ]\kern -.4em]$}}

\newcommand{\nl}{\medskip \noindent }

\newcommand{\N}{\mathbb{N}}

\newcommand{\Z }{\mathbb{Z}}
\newcommand{\Q}{\mathbb{Q}}

\newcommand \str [1] {$\langle  {#1} \rangle$}

\newcommand{\cor}[1]{{\color{red} {#1}}}

\newcommand{\li}[1]{\inf({#1})}
\newcommand{\ls}[1]{\sup({#1})}

\newcommand{\aag}{{\cal A}^{(+)}}  %sous-automate gauche
\newcommand{\aad}{{\cal A}^{(-)}}  %sous-automate droit

\usetikzlibrary{arrows,automata,plotmarks,decorations,positioning}
\tikzstyle{post}=[->,shorten >=1pt,>=stealth']
\tikzstyle{every initial by arrow}=[initial text={},initial distance=1em,post]
\tikzstyle{transition}= [post,shorten >=1pt,node distance=2cm, inner sep=2pt,bend angle=20]

\newlength{\transwidth}
\newcommand{\trans}[1]{\settowidth{\transwidth}{$\;_{#1}\;$}
\stackrel{#1}{\overrightarrow{\rule{\transwidth}{0ex}}}}

\begin{document}

\title{ 
Synchronous orders on the set of integers
}

\author{ Christian Choffrut\\
IRIF (UMR 8243), \\
CNRS and Universit\'e Paris Cit\'{e},  
France\\
\tt{Christian.Choffrut@irif.fr}
}
\date{}
\maketitle

\tableofcontents

\bigskip 
{\cor{A late bibliographical search showed me that most of the material presented is not new. 
The characterization of linear synchronous orderings on $\N$ is mentioned in \cite[Thm 4]{LM2009}, which  refers to \cite{KR2001} where it is implicit in Theorem 9. In \cite[Thm 4.1]{LM2011} it is shown that  
the isomorphism 
of two linear synchronous orders on $\N$  is decidable in linear time in the sizes of the input automata.
These results are obtained by considerations on the   structures of general automatic graphs by introducing the notion
of graph unwinding \cite[page 473]{KR2001}.

End of  july 2023 Jeff Shallit asked me if I knew whether of not the following problem is decidable: assuming that a given synchronous automaton
recognizes an ordering, does this ordering have infinite chains? infinite antichains? and questions of the like. I suspected that this is not the case but I started to work in the special case where the alphabet has a unique letter. I could prove that these questions are decidable, I characterized the orders that are linear and showed that the equivalence of two linear orders is decibable. I posted it on arxiv on  september 19.  Then I thought that I should enrich my introduction \ldots and found out that these were old results. I did not change the rest of my manuscript except for the bibliography. }.

\begin{abstract} A binary relation over a free monoid is synchronous if it can be recognized by a synchronous automaton that reads 
its two tapes simultaneously. We consider the case where the free monoid is generated by a single element (which makes it isomorphic to the additive monoid of integers) and where the binary relation recognized is a strict order. Our main results are: given such an automaton it is possible to determine whether or not is has infinite chains or antichains; we characterize the orders that are linear; given two linear synchronous orders we show how  to determine whether or not they are equivalent.
\end{abstract}

%%%%%%%%%%%%%%%%%%%%%%%%%%%%%%%%%%%%%%%%%%%
%%%%%%%%%%%%%%%%%%%%%%%%%%%%%%%%%%%%%%%%%%%
\section{Introduction}
%%%%%%%%%%%%%%%%%%%%%%%%%%%%%%%%%%%%%%%%%%%
%%%%%%%%%%%%%%%%%%%%%%%%%%%%%%%%%%%%%%%%%%%

\bigskip 
Let $\+A$ be a structure over some domain $D$ with a collection of relations. It is 
\emph{automatic} if there is some finite alphabet such that via an appropriate encoding,  $D$ maps into 
 a regular language. Furthermore each $n$-ary relation can be encoded into a synchronous finite automata
which are $n$-tape automata reading the $n$ tapes simultaneously. Consequently,  the first order  theory of these structures 
is decidable. There is a general agreement to date the coining of the term to \cite{hodgson1976}
but the systematic study of discrete groups via automatic and semiautomatic structures \cite{epstein1992} is often ignored.
Most of the  literature in this area consists of inquiring which structures have or do not have an automatic presentation 
(see, e.g., \cite{gradel20} for a good account).

Here our purpose is opposite. Instead of starting with structures we start with automata. More specifically,
considering synchronous automata over a unary alphabet, 
 i.e., over  the set of natural integers $\N$, what kind of order structures can they possibly represent? 
We show that given such an automaton we can determine whether or not the order has infinite chains or antichains. We are able to characterize
the linear orders that are representable and show that given two synchronous automata representing linear orders, it is decidable whether or not 
these orders are equivalent. These questions can be posed for nonunary synchronous automata but I am not aware of any published results in the
general case.  Concerning nonunary alphabets, the situation is to be compared with that of relations defined by 
 $n$-tape automata which process $n$ tapes from left to right but which are not constrained to read  them simultaneously. The family thus obtained,  the \emph{rational relations}  in the terminology of \cite{eilenberg1974} or the relation defined by generalized automata in \cite{EM1965}, see also \cite{saka2009},  is much richer than the family of synchronous
relations. In particular, the most basic properties such as reflexivity, antisymmetry and transitivity are undecidable for this class, \cite[pages 56-57]{johnson1986}. This makes the decidability of the questions tackled in this paper for synchronous relations over nonunary alphabets more challenging.

%%%%%%%%%%%%%%%%%%%%%%%%%%%%%%%%%%%%%%%%%%%
%%%%%%%%%%%%%%%%%%%%%%%%%%%%%%%%%%%%%%%%%%%
\section{Preliminaries}
%%%%%%%%%%%%%%%%%%%%%%%%%%%%%%%%%%%%%%%%%%%
%%%%%%%%%%%%%%%%%%%%%%%%%%%%%%%%%%%%%%%%%%%

%%%%%%%%%%%%%%%%%%%%%%%%%%%%%%%%%%%%%%%%%%%
\subsection{Synchronous relations}
%%%%%%%%%%%%%%%%%%%%%%%%%%%%%%%%%%%%%%%%%%%

With every $n$-tuple of words $(x_{1},\ldots, x_{n})$ of a finite alphabet $\Sigma$ we associate the $n$-tuple
obtained by padding to the right of each component as few occurrences of a new symbol $\#$ as possible in such a way that all components have the same length, e.g., $(a,aba,bb)^{\#}= (a\#\#,aba, bb\#)$. Every padded $n$-tuple can be considered unambiguously as an element of the free monoid generated by the finite alphabet
$\Delta= (\Sigma \cup \{\#\})^{n} \setminus \{(\#)^{n}\}$.
E.g., $(a,aba,bb)^{\#}= (a\#\#,aba, bb\#)=(a,a, b)(\#,b, b)(\#,a, \#) $.
Given an $n$-ary relation $R\subseteq \Sigma^{n}$, we let $R^{\#}$ denote the set 
$\{(x_{1},\ldots, x_{n})^{\#} \mid (x_{1},\ldots, x_{n})\in R\}$. Then $R\subseteq (\Sigma^*)^{n}$ is \emph{synchronous} if there exists a finite automaton on 
the alphabet $\Delta$ which recognizes $R^{\#}$. This notion was introduced by Elgot and Mezei 
with the terminology of  ``FAD'' relation.
It was proved that these relations form a Boolean algebra, that their class is closed under composition of relations and 
under projection
and that the emptiness problem  is decidable  \cite[page 49]{EM1965}.

It is proved in \cite{EES1969} that the set of synchronous relations on a nonunary alphabet
is the set of $n$-tuples defined in the first order logic of $\Sigma^*$ with the signature consisting of 
the prefix relation (the string $x$ is a prefix of the string $y$), the equal length relation ($x$ and $y$ have equal length)
and last letter unary relation (some $a\in \Sigma$ is the last letter of the string $x$). However this logic fails to capture all synchronous relations in the case of a unary alphabet (see paragraph \ref{ss:synchronous-unary-alphabet} for a suitable logic in this case).

%%%%%%%%%%%%%%%%%%%%%%%%%%%%%%%%%%%%%%%%%%%
\subsection{Synchronous automata on a unary alphabet}
\label{ss:synchronous-unary-alphabet}
%%%%%%%%%%%%%%%%%%%%%%%%%%%%%%%%%%%%%%%%%%%

The free monoid over a unary alphabet is commutative and can be identified with $\N$. The concatenation is written additively. Instead of the ugly padding symbol, we view the $n$-vectors in $\N^n$ differently. The \emph{support}
of a vector $x\in \N^n$ is the subset of indices $0<i\leq n$ such that $x_{i}\not=0$. 
With all subsets $\emptyset \not= I\subseteq \{1,\ldots, n\}$ we associate the vector $e_{I}$ whose $i$-th component is equal to $1$ if
$i\in I$ and to $0$ otherwise. Then every nonzero vector can be written uniquely as a sum 
$e_{I_{1}} + \cdots + e_{I_{k}}$ where $I_{1}\supseteq  \cdots \supseteq I_{k}$. 
Reinterpreting the use of the $\#$ symbol in this particular case, a synchronous automaton on  $\N^n$ is a finite automaton on the finite alphabet $\{e_{I}\mid \emptyset \not= I\subseteq \{1,\ldots, n\}\}$  with the condition that if $e_{I}$ and 
$e_{J}$ label two cosecutive transitions, then $I\supseteq J$.

\medskip A different proof of the  following can be found in  \cite{Peladeau92}
where this logic is called  \emph{modular logic}. Further results on this logic can be found in \cite{choffrut08}

\begin{proposition}
\label{pr:synchronous-relations}
A relation $R\subseteq \N^n$ is synchronous if and only if it can be defined in the first order logic
of the structure
\str{\N;  (x-y\in L)_{L\text{ regular}}}
\end{proposition}

\begin{proof} It is routine to check that all the primitive predicates define synchronous relations.
Furthermore, the synchronous relations form a Boolean algebra and are closed under composition and projection,
\cite{EM1965}, so all definable relations are synchronous. We prove the converse
by showing that the running of the automaton can be defined by a formula.

The set $E=\{e_{I}\mid I \subseteq  \{1,\ldots, n\} \}$ is provided with 
 the partial ordering $e_{I}\geq e_{J}$ if $J\subseteq I$.
Now every vector in $\N^{n}$ can be uniquely expressed as $\alpha_{1} e_{I_{1}} +\cdots + \alpha_{r} e_{I_{r}}$
where $e_{I_{1}} > \cdots > e_{I_{r}}$,  $\alpha_{1}, \cdots,  \alpha_{r} \in \N\setminus \{0\}$.

The set of states $Q$ of a synchronous automaton is a disjoint union of subsets 
$ Q_{I}$, $I\subseteq \{1,\ldots, n\}$ where a state $q$ belongs to $Q_{I}$
if and only if  it is the target of  a transition labeled by $e_{I}$. 
 A generic path is of the form 
\begin{equation}
\label{eq:paths}
 q_{0} \trans{\alpha_{1} e_{I_{1}} }  q_{1} \trans{\alpha_{2} e_{I_{2}} } \cdots  q_{r-1}  \trans{\alpha_{r} e_{I_{r}} }   q_{r} 
 \end{equation}
 and the relation recognized is the union over all sequences 
 $\{1, \ldots, n\}\supset I_{1} \supset I_{r-1}\supset I_{r}$ and over all sequences of states 
$ q_{0},   q_{1},  \ldots,    q_{r} $ of the labels $\alpha_{1} e_{I_{1}} +  \cdots  + \alpha_{r} e_{I_{r}} $.
 We let $L_{k}$ denote the (regular) set of lengths of all paths from $q_{k-1}$ to $q_{k}$. The set 
 of labels of a path such as \ref{eq:paths} can be expressed by the following formula
$\bigwedge_{1\leq k\leq r} \phi_{k}$ with 
$$
\phi_{k}=(\bigwedge_{i,j\in I_{k}} x_{k, i}=x_{k, j}) \wedge (\bigwedge_{i\in I_{k}} x_{k,i}-0\in L_{k})
$$
if $r=1$. Otherwise, with the convention $I_{r+1}=\emptyset$ 
$$
\phi_{k}=(\bigwedge_{i,j\in I_{k}\setminus I_{k+1}} x_{k, i}=x_{k, j}) \wedge (\bigwedge_{i\in I_{k}} x_{k,i}-x_{k-1,i}\in L_{k})
$$
where $x_{k-1,1}$ is interpreted as $0$ if $k=1$. 

\end{proof}

%%%%%%%%%%%%%%%%%%%%%%%%%%%%%%%%%%%%%%%%%%%
\subsection{The binary case}
\label{ss:binary-case}
%%%%%%%%%%%%%%%%%%%%%%%%%%%%%%%%%%%%%%%%%%%

In this paragraph, we describe explicitly the natural decomposition of the synchronous automata on $\N\times \N$
and fix some notations so as to be able to work more easily on binary relations which is the main purpose of this paper.

Given a symbol $a$, an $a$-deterministic automaton over the alphabet  $\{a\}$ consists of $n$ states $q_{0}, \ldots, q_{n-1}$ along with 
 $n-1$ transitions of the form $q_{i} \trans{a} q_{i+1}$ for $0\leq i<n-1$ and 
$q_{n-1} \trans{a} q_{t}$   for some $\leq t<n-1$.  The integer $p=ln-t$ is the \emph{period} and $t$ the \emph{transient}.
The state $q_{0}$ is the inital state of the automaton. A state $q_{i}$  \emph{is transient} if $i<t$ and \emph{periodic} otherwise.

 The following proposition is trivial and could serve as a definition of binary synchronous automata. 

\begin{proposition}
\label{pr:synchronous-automaton}
A binary synchronous automaton ${\cal A}$ consists of an $(1,1)$-deterministic automaton ${\cal B}$ 
whose state  set is $\{q_{0}, \ldots, q_{n-1}\}$ and for $0\leq i<n$ an
 $(1,0)$- resp.  $(0,1)$- deterministic automaton   $\aag_{i}$  resp. $\aad_{i}$
  satisfying the following conditions

\begin{itemize}
\item the initial state of $\aag_{i}$  and  $\aad_{i}$   is $q_{i}$.

\item the set of states of ${\cal B}$ and   $\aag_{i}$ for $0\leq i<n$ are disjoint except for the initial state of $\aag_{i}$.
The same holds with  ${\cal B}$ and   $\aad_{i}$ for $0\leq i<n$.
The set of states of   $\aag_{i}$ and   $\aad_{j}$ for $i\not=j$ are disjoint.

\item a state of $\+A$ is final if and only if it is a final state of some $\aag_{i}$  or $\aad_{j}$.
 \end{itemize}
\end{proposition} 
 
 The relation $R\subseteq \N\times \N$ \emph{defined} or \emph{recogn-ized} by $\+A$ is the set of pairs $(k,\ell)$ which label a path from $q_{0}$ to a final state of $\aag_{i}$  or $\aad_{j}$. 
 \begin{example}
 
 The automata below recognize the linear order $\cdots < 2<1<0$ and
$0<1<2\cdots $

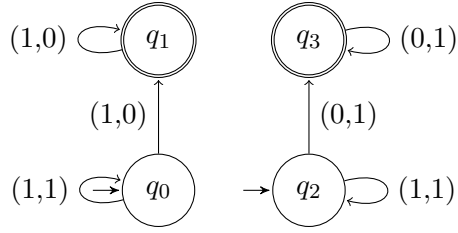
\begin{figure}[H]
\begin{center}
 \usetikzlibrary {automata,positioning}
\begin{tikzpicture}[shorten >=1pt,node distance=2cm,on grid,auto]

  \node[state,initial]  (q_0)                      {$q_0$};
\node[state,accepting]           (q_1) [above =of q_0] {$q_1$};
\node[state,initial]           (q_2) [right =of q_0] {$q_2$};
 \node[state,accepting]         (q_3) [above =of q_2] {$q_3$};
 \path[->] (q_0) edge              node        {(1,0)} (q_1)
  edge [loop left] node        {(1,1)} ()
 (q_1)  edge [loop left] node        {(1,0)} ();

  \path[->] (q_2) edge   [right]           node        {(0,1)} (q_3)
  edge [loop right] node        {(1,1)} ()
 (q_3)  edge [loop right] node        {(0,1)} ();

\end{tikzpicture}
\label{fig:omegastar}
\caption{the relations $\cdots < 2<1<0$. and $0<1<2\cdots $}
\end{center}
\end{figure}
The next automaton recognizes the linear orders $2<1<0$ (three elements)
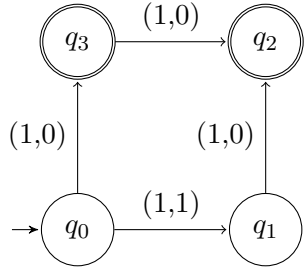
\begin{figure}[H]
\begin{center}
 \usetikzlibrary {automata,positioning}
\begin{tikzpicture}[shorten >=1pt,node distance=2.5cm,on grid,auto]

  \node[state,initial]  (q_0)                      {$q_0$};
\node[state]           (q_1) [right =of q_0] {$q_1$};
\node[state,accepting]           (q_3) [above  =of q_0] {$q_3$};
\node[state,accepting]           (q_2) [above  =of q_1] {$q_2$};

 \path[->] (q_0) edge              node        {(1,1)} (q_1)
 (q_0) edge                node        {(1,0)} (q_3)
(q_3) edge                node        {(1,0)} (q_2)
% (q_0) edge       [bend left]              node        {(1,0)} (q_2)
(q_1) edge            node        {(1,0)} (q_2) ;
\end{tikzpicture}
\label{fig:two}
\caption{the relation $2<1<0$}
\end{center}
\end{figure}

 \end{example}

 Strictly speaking, the labels of the nonempty paths in $\+B$ are pairs of the form $(k,k)$ with $k>0$. By convention we identify 
 $(k,k)$ with  the integer $k$. Similarly, the labels of a path in $\aag_{i}$ (resp. $\aad_{i}$
 are of the form  $(k,0)$ (resp. $(0, k)$)
 which we identify with the integer $k$. As a result, an integer $k>0$ is interpreted as  $(k,k)$ in ${\cal B}$,
  $(k,0)$  in $\aag_{i}$ and $(0, k)$  in   $\aad_{i}$.
 We let $\lambda_{i}$ and $\rho_{i}$ denote the transitions on the subautomata $\aag_{i}$ and $\aad_{i}$. Thus we write
 $\lambda_{i}(q_{i},k)=r$ and $\theta_{i}(q_{i}, \ell)=s$ with the obvious meaning.
 The transition on ${\cal B}$ is simply denoted $q\cdot k$ for all integers $k$ and all states $q$ of ${\cal B}$.
 By abuse of notation, for all $i\geq n$ we let $q_{i}$ be the states $q_{j}$ where $n-p\leq j<n$ and 
 $j=i\bmod p$.
  With these notations we have 
  $R(k,\ell)$  if and only if

  \begin{equation}
  \label{eq:fundamental}
  \begin{array}{l}
\text{if  }   k<\ell \text{ and }  q_{0}\cdot  k= q_{i} \text{ then }  \theta_{i}(q_{i}, \ell-k) \text{ is final in  }  \aad_{i}\\
\text{if  }    k>\ell \text{ and }    q_{0} \cdot  \ell= q_{i} \text{ then }  \lambda_{i}(q_{i}, k-\ell) \text{ is final in  }  \aag_{i}
 \end{array}
  \end{equation}

%%%%%%%%%%%%%%%%%%%%%%%%%%%%%%%%%%%%%%%%%%%
%%%%%%%%%%%%%%%%%%%%%%%%%%%%%%%%%%%%%%%%%%%
 \section{Orders}
%%%%%%%%%%%%%%%%%%%%%%%%%%%%%%%%%%%%%%%%%%%
 %%%%%%%%%%%%%%%%%%%%%%%%%%%%%%%%%%%%%%%%%%%

 %%%%%%%%%%%%%%%%%%%%%%%%%%%%%%%%%%%%%%%%%%%
 \subsection{General definitions}
%%%%%%%%%%%%%%%%%%%%%%%%%%%%%%%%%%%%%%%%%%%

By an order on a set $X$ we mean  a strict partial order, i.e., a 
binary relation which is transitive and has no loop, i.e., that satisfy the axioms.
$$ \forall x,y,z\ R(x,y)\wedge R(y,z)\rightarrow R(x,z)\quad  \neg \exists x,y\ R(x,y)\wedge R(y,x)
$$
The \emph{support} of $R$, denoted $\emph{supp}(R)$ is the subset $\{x\in X\mid \exists y\  (x,y)\in R \vee (y,x)\in R\}$.
The order is \emph{complete} if the support of $R$ is $X$.
The relation  is \emph{linear} if for all $x,y$ in the support of $X$ either $(x,y)\in R$ or $(y,x)\in R$ holds. E.g., 
the relation $(2n,2m)$ with $n<m$ defines a linear order on $\N$ but it is not complete.

Two ordered sets $X$ and $Y$ are said to have the same \emph{order type} 
 if  there exists a bijection $f:X\rightarrow Y$ 
 such that  $f$ and its its inverse are monotonic.  The order types 
of $\N$, $-\N$, $\Z$ and $\Q$
are denoted $\omega$, $\omega^*$, $\zeta$ and  $\eta$ respectively. Given an integer $n\in \N$ the linear order type 
of $\{1, 2, \ldots, n\}$ is denoted  $\mathbf{n}$. The reader is referred to the handbook  \cite{rosenstein} for
 an introduction to linear orders. 
 The \emph{sum}  $R+S$ of two orders of disjoint supports is the order defined by 
\begin{equation}
\label{eq:sum-of-synchronous}
 R\cup S \cup \text{supp}(R)\times \text{supp}(S)
\end{equation}
The sum of two order types $\rho$ and $\sigma$ is the order type $\rho +\sigma$ of the sum 
$R+S $ for any $R$ of order type $\rho$ and $S$ of order type $\sigma$
of disjoint supports.
The \emph{inverse} of $R$ is the relation $R^{-1}=\{(y,x) \mid (x,y)\in R\}$.

An order on $\N$ is \emph{synchronous} if there exists a synchronous automaton on $\N$ such that 
$R$ is the set recognized by the automaton.

\emph{Henceforth, all synchronous orders are orders on $\N$}.

%%%%%%%%%%%%%%%%%%%%%%%%%%%%%%%%%%%%%%%%%%%
\subsection{Poor linear orderings}
%%%%%%%%%%%%%%%%%%%%%%%%%%%%%%%%%%%%%%%%%%%

We find it convenient to say that a linear order type is \emph{poor} if it is 
a finite sum of order types $\omega$, $\omega^*$ and $\mathbf{n}$ for some 
$n\in \N$. 
\begin{equation}
\label{eq:poor-sum}
\chi_{1}+ \cdots + \chi_{n} \quad \chi_{i}=\omega \text{ or } \omega^* \text{ or } \mathbf{n} \text{ for some } n<\omega
\end{equation}
This sum  is 
\emph{reduced} if it is equal to $\mathbf{0}$ or otherwise if it contains no subsums of the form
$\mathbf{0}$,  $\mathbf{n}+ \mathbf{m}$,  $\mathbf{n}+ \omega= \omega$ or 
 $\omega^* + \mathbf{n} =  \omega^*$.

\begin{proposition}
\label{pr:reduced-expression}
Every poor linear order is equivalent to a unique reduced sum.
\end{proposition}

\begin{proof}  Consider two reduced sums $S_{1}$ and $S_{2}$ defining the same poor order. 
We  show that they are equal by induction on the maximum of number of summands in $S_{1}$ and $S_{2}$
and we assume that none of them is reduced to $\mathbf{0}$.
If the order has no minimum element then $S_{1}= \omega^* + T_{1} $ and $S_{2}= \omega^* + T_{2} $
then $T_{1}$ and 
$T_{2}$ are equivalent and we are done. If  we have
%

%\break

\begin{eqnarray}
\label{eq:unicity}
\nonumber
S_{1}= \mathbf{n}_{1} + \chi_{1} + T_{1} \\
\nonumber
S_{2}= \mathbf{n}_{2} + \chi_{2} + T_{2}
\end{eqnarray}
with $n_{1}, n_{2}. \not=0$ then $ \chi_{1} =\chi_{2}= \omega^*$ which implies $n_{1}= n_{2}$, i.e., $\chi_{1} + T_{1}$ 
and $\chi_{2} +T_{2}$ are equivalent and we are done.  So  we assume without loss of generality  $n_{2}>0$
$$
S_{1}= \chi_{1} + T_{1} \quad 
S_{2}= \mathbf{n}_{2} + \chi_{2} + T_{2}
$$
Then $\chi_{1}=\omega$ because the order has a minimal element 
and $\chi_{2}=\omega^*$ but this is impossible because  $\omega$ and $\mathbf{n}_{2} + \omega^*$ 
are incomparable.

 \end{proof}
 
 \medskip 
For all subsets $A$ of a linear order we let  $ \li{A} $ denote the greatest lower bound and 
 $\ls{A} $ the least upperbound of $A$ when they exist.  For two subsets $A,B$  we write  $A\prec B$ if for all $a\in A$ and $b\in B$ it holds 
$R(a,b)$.

\begin{proposition} 
\label{pr:charac-poor-linear-orders}
The order type of a linear relation is poor 
if and only if its support is a finite union of singletons,  countable ascending and countable descending chains. 
\end{proposition}

\begin{proof} The condition is clearly necessary. We prove that it is sufficient by induction on the number of  
chains and singletons.

\medskip

We let  $I(X)$ denote the minimal interval conatining $X$.
If the set consists of a unique singleton or of a unique ascending or descending chain then we are done.
Assume that we have proved that a linear set which 
a finite union of singletons and infinite ascending and descending chains 
is a disjoint union  $ A=\bigcup^{r}_{i=1} A_{i} $ where for all $0<i<r-1$
we have $I(A_{i} )\prec I(A_{i+1} )$. %
Since poor linear orderings are closed under inverse ordering
 it suffices to prove that the  union of $A$ with  a subset $B$ which is either a singleton or 
 of order type $\omega$, 
has an expression of the form \ref{eq:poor-sum}. 
We consider the case where $B$ is a singleton.
If it belongs to  
some $I(A_{i})$ then 
 $A_{i}$ is replaced by  $A'_{i}=A_{i}\cup B$. The order type is  
unchanged if $A_{i}$ is of type $\omega$ or $\omega^*$ or  if $B\subseteq A_{i}$ and
otherwise  it is changed to $\mathbf{n+1}$ if $A_{i}$
is of type  $\mathbf{n}$ with $n>1$. If $B$ belongs to no such interval then  
$B$ is inserted between $A_{i}$ and $A_{i+1}$ or before $A_{1}$ or after $A_{r}$ accordingly.

Now we consider the case where $B$ is of type $\omega$. Every interval contains either 
a finite subset of $B$ or the complement of a finite prefix. Because of the previous
 consideration we may  assume that $B$ is included in some $I(A_{i})$
or in some interval separating the $A_{i}$'s.
If  $B\prec A_{1}$ or  $A_{r} \prec B$ or if $B$ is contained in the interval separating 
$A_{i}$ and   $A_{i+1} $  for some $i=1, \ldots, r-1$
then the disjoint union is obtained by inserting $B$ either before $A_{1}$ of after $A_{r}$
or between $A_{i}$ and $A_{i+1}$. 
 
Assume it holds   $B\subseteq A_{i}$
 which implies in particular that $A_{i}$ is of type $\omega$ or $\omega^*$. If $A_{i}$ is of type $\omega$
 and  $\ls{B} = \ls{A_{i}}$ then $A_{i}\cup B$ is of type $\omega$ otherwise it is of type $\omega\cdot 2$ because
$$
A_{i} \cup B =  \big(B\cup (A_{i}\cap [\li{A_{i}}, \ls{B}])\big) \cup (A_{i} \cap ]\ls{B}, \ls{A_{i}}[) %& \omega \text{ type}%\\
$$

There remains the case where   $A_{i}$ is of type $\omega^*$. Then we have $\li{A_{i}}<\li{B} <\ls{B}< \ls{A_{i}}$
which implies 
$$
\begin{array}{l}
A_{i}\cup B = A_{i,1} \cup A_{i,2} \cup A_{i,3} \quad A_{i,1} \prec A_{i,2}  \prec  A_{i,3} \\
A_{i,1}  =(]\li{A_{i}},\li{B}[\cap A_{i})\\
A_{i,2} = ([\li{B},\ls{B}[ \cap (B\cup A_{i}))\\
A_{i,3}  = [\ls{B}, \ls{A_{i}}]\cap A_{i}
\end{array}
$$
whose order type is $\omega^* + \omega +\mathbf{n}$.

\end{proof}
%
%%%%%%%%%%%%%%%%%%%%%%%%%%%%%%%%%%%%%%%%%%%
%%%%%%%%%%%%%%%%%%%%%%%%%%%%%%%%%%%%%%%%%%%
\section{Synchronous orders on $\N$}
%%%%%%%%%%%%%%%%%%%%%%%%%%%%%%%%%%%%%%%%%%%
%%%%%%%%%%%%%%%%%%%%%%%%%%%%%%%%%%%%%%%%%%%

We recall that unless otherwise stated we deal with synchronous relations $R\subseteq \N\times \N$ uniquely. 
We make additional technical assumptions to Proposition  \ref{pr:synchronous-automaton}
which simplify the proofs without losing the generality
of the results. An automaton is \emph{normal} if it satisfies the following conditions
(with the notations of Proposition  \ref{pr:synchronous-automaton})

\begin{itemize}
\item The subautomaton $\cal B$ and all automata $\aag_{i}$  and. $\aad_{j}$  have the same period $p$ which is 
 greater than their transients.

\item All pairs $({k}, {\ell})$ define exactly one path  in 
$\+A$, whether successful or not. 

 \end{itemize}

\begin{proposition}
\label{pr:normal-automaton}
A synchronous relation $R$ can be realized by a normal automaton.
Let $t\leq \alpha < \beta<n$ with the notations of  Proposition  \ref{pr:synchronous-automaton}.
For $R^{\epsilon}$ equal to $R$ or its inverse $R^{-1}$  we have
 \begin{eqnarray}
\label{eq:a2b}
(\alpha +2p ,\beta)\in R^{\epsilon} \Leftrightarrow \forall k\geq 2\  (\alpha + k p, \beta)\in R^{\epsilon}\\
\label{eq:ab}
(\alpha,  \beta + p)\in R^{\epsilon} \Leftrightarrow \forall k\geq 1\ (\alpha,  \beta + k p)\in R^{\epsilon}\\
\label{eq:deflating}
\forall k< \ell  \quad 
\label{eq:deflating}( \alpha +kp, \beta +\ell p)\in R^{\epsilon} \Rightarrow (\alpha,  \beta + (\ell-k)p)\in R^{\epsilon}
\end{eqnarray}
\end{proposition}
 
\begin{proof} It suffices to consider the case $R^{\epsilon} =R$. We set $\alpha = t+r, \beta =t + s, 0\leq r<s<p$ 
and we use the statement \ref{eq:fundamental}.

\nl Implication \ref{eq:a2b}. We have  $\alpha +2p -\beta = p+(s-r)>p$, 
thus the  pair $(\alpha +2p ,\beta)$ takes $q_{0}$ to $\lambda_{\beta}(q_{\beta},p+(s-r)$ which is a periodic state of 
$\aag_{\beta}$, implying that for $k\geq 2$,  $(\alpha +kp ,\beta)$ takes $q_{0}$ to $\lambda_{\beta}(q_{\beta},2p+(s-r))
=\lambda_{\beta}(q_{\beta},p+(s-r))$.%$\theta_. \lambda_$

\nl Implication \ref{eq:ab}. We have $\beta +p -\alpha= s-r+p>p$  
thus the  pair $(\alpha,  \beta + p)$ takes $q_{0}$ to $\theta_{\alpha}(q_{\alpha},p+(s-r)$ which is a periodic state of 
$\aad_{\alpha}$, implying that for $k\geq 1$,  $(\alpha, \beta +kp)$ takes $q_{0}$ to $\theta_{\alpha}(q_{\alpha},s-r+kp)
=\theta_{\alpha}(q_{\alpha},s-r+p)$.

\nl Implication \ref{eq:deflating}. We have $q_{0}\cdot \alpha = q_{0}\cdot  k \alpha$  
and  $\beta +\ell p -(\alpha +kp)= ( \ell -k)p +s-r>p$. Now, $\theta_{\alpha}(\beta,  u p)$ is a periodic state of 
$\aad_{\alpha}$ for all $u\geq 1$. Thus  $(\alpha,  \beta + (\ell-k)p)$ takes $q_{0}$ to 
$\theta_{\alpha}(q_{\alpha}, \beta +(\ell-k) p)$ and $(\alpha, \beta +\ell p)$ takes $q_{0}$ 
to %$(\alpha +k p,  \beta + \ell p))$ takes 
 $\theta_{\alpha}(q_{\alpha}, \beta +\ell p)= \theta_{\alpha}(q_{\alpha}, \beta +(\ell-k) p)$. 

\end{proof}
 
%%%%%%%%%%%%%%%%%%%%%%%%%%%%%%%%%%%%%%%%%%%
\subsection{Composition}
%%%%%%%%%%%%%%%%%%%%%%%%%%%%%%%%%%%%%%%%%%%

The following is a simple way to define a synchronous relation equivalent to a given synchronous relation
of arbitrary arity (the arity is arbitrary and the relation is not necessarily an order).

\begin{lemma}
\label{le:clone-of-relations}
If $R\subseteq \N^n$ is synchronous then for all integers $m>r\geq 0$ the relation $R_{m,r}$ defined by 
$(m x_{1} +r, \ldots, m x_{n} +r)\in R_{m,r}$ if and only if 
$(x_{1}, \ldots, x_{n})\in R$
is also synchronous.

If $n=2$  and $R$ defines an ordering, then $R_{m,r}$ defines an equivalent ordering.
\end{lemma}

\begin{proof}
Let $\cal A$ be a synchronous automaton recognizing $R$. The relation $R_{m,0}$ is recognized by the automaton obtained from $\cal A$ by replacing  each transition $q \trans{x} q'$ where $x\in \{0,1\}^{n}$ by a sequence of $m$ transitions 
$$q \trans{x} s_{1} \cdots  s_{m-1} \trans{x} q'
$$
where the $s_{i}$ are fresh state symbols.
If $\phi(x_{1}, \ldots, x_{n})$ defines the relation $R_{m,0}$ then the relation $R_{m,r}$
is defined by the formula
$$
\exists y_{1}, \ldots, y_{n}\ (\bigwedge^{n}_{i=1} x_{i}=y_{i} +r \ \wedge \phi(y_{1}, \ldots, y_{n}))
$$

\end{proof}

We can construct new relations with the help of  the previous lemma.

\begin{proposition}
\label{pr:sum-of-two}
If $R, R' \subseteq \N^2$ are two synchronous orders of  types $\rho, \rho'$ 
then the sum $R + R' $ is a synchronous relation  of order type  $\rho+ \rho'$.
 \end{proposition}

\begin{proof}
It suffices to verify that $R + R'$  is synchronous, but this results from the expression
$$
R+R'= R\cup R' \cup \text{supp}(R)\times \text{supp}(R')
$$

 \end{proof}

%%%%%%%%%%%%%%%%%%%%%%%%%%%%%%%%%%%%%%%%%%%
  \subsection{Characterization of synchronous linear orders}
 %%%%%%%%%%%%%%%%%%%%%%%%%%%%%%%%%%%%%%%%%%%

It is clear that $\omega, \omega^*$ and $\mathbf{n}$ are synchronous (e.g., see  Figure \ref{fig:omegastar}
for an order type $\omega^*$ and $\omega$ and Figure \ref{fig:two}. for an order of type 
$\mathbf{3}$ ). By Proposition \ref{pr:sum-of-two} 
all  linear relations of poor type 
are synchronous. The converse holds but we need two previous elementary results. 

\begin{lemma}
\label{le:linear-on-regular}  Let $E\subseteq \N$ be a regular subset. The linear relation  $\{(a,b)\mid (a<b, a,b\in E\}$ is synchronous. Similarly for the linear order  $\{(a,b)\mid (a>b, a,b\in E\}$.
\end{lemma}

\begin{proof}
 Using Proposition \ref{pr:synchronous-relations},  the trace on $E$ of the natural order over $\N$ is defined by the  formula
$$
x-0\in E \wedge y-0\in E \wedge x-y>0
$$
\end{proof}

\begin{corollary}
\label{cor:infinite-complement}  Let $R$ be  synchronous and assume that 
 the complement $E$ of its support is infinite. The exists a synchronous relation $S$ of support $E$ 
and  of order type $\omega$ (resp. $\omega^*$) such that $R+S$ is complete.
 If $\tau$ is the order type of $R$, then the order type of 
  $R+S$ is $\tau+\omega$.   
\end{corollary}

\begin{lemma}
\label{le:finite-complement}  Let $R$ be a  synchronous
relation. If $\text{supp}(R)$ has a finite complement, there exists a  complete
synchronous   order which has the same order type as $R$.
\end{lemma}

\begin{proof}
Indeed, assume $a\not\in \text{supp}(R)$.
Define $f:\N \rightarrow \N$  by $f(k)=k$ is $k<a$ and 
$f(k)= k-1$ if $k>a$. Consider the relation $R'=\{(f(k), f(\ell)) \mid (k,\ell)\in R\}\}$. The relation is an order and it is synchronous because it is
the composition of three synchronous  relations $f^{-1} \circ R\circ f$. Furthermore $f$ and $f^{-1}$ ar monotone.
If the complement of the support of $R$  contains $b$ elements, it suffices to apply this construction $b$ times.

\end{proof}

\begin{theorem}
\label{th:char-linear}
A countable linear relation on $\N$ is synchronous if and only if it is of poor order   type, e.g., of the form   \ref{eq:poor-sum}.

 \end{theorem}

 \begin{proof} The condition is sufficient by Proposition \ref{pr:sum-of-two}.
We prove that it is necessary.  By lemma \ref{le:finite-complement}    and corollary \ref{cor:infinite-complement}  
we may assume that the relation is  complete since
for any order type $\tau$,  $\tau$ is a poor linear order if and only if $\tau+\omega$ is a poor linear order.

  \medskip We write $n\prec m$ if $R(n,m)$ holds and reserve the expression $n<m$ for the usual order in the integers. 
  The expression $n\succ m$ is equivalent to $m\prec n$.   
   For some  $t\leq  \alpha <t+ p$ consider the set of inputs greater than 
  or  equal to $t$ that are congruent to $\alpha$ modulo $p$.
I claim that we  have 
\begin{eqnarray}
%\label{ascending}
 \alpha\prec   \alpha +p \prec   \cdots \prec  \alpha + kp \prec \cdots
\label{ascending}
\text{ or }\\
\label{descending}
 \alpha \succ  \alpha +p \succ   \cdots \succ  \alpha + kp \succ\cdots
\end{eqnarray}
 Since the order is total we have  $ \alpha \prec \alpha +p $
  or  $ \alpha \succ \alpha +p $. In the first case we have $q_{0}\cdot \alpha= q_{i}$ and $\theta_{i}(q_{i},p)= r$. But then 
 $q_{0}\cdot (\alpha+p)= q_{i}$ and $\theta_{i}(q_{i},p)= r$, i.e., $ \alpha +p\prec \alpha + 2p $. 
 More generally we have $ \alpha +k p\prec \alpha + (k+1) p $ which proves the claim. The second case can be treated similarly.
 Since the relation is a finite union of singletons and countable chains as above we may conclude by applying Proposition
 \ref{pr:charac-poor-linear-orders}.
 
 %%with the linear order  on $(\{0\}\times 2\Z) \cup (\{1\}\times 2\Z +1)$ defined $(a,n<(b,m)$ if $n<m$.}

\end{proof}

%%%%%%%%%%%%%%%%%%%%%%%%%%%%%%%%%%%%%%%%%%%
 \subsection{Decision issues of on unary synchronous orders.}
 %%%%%%%%%%%%%%%%%%%%%%%%%%%%%%%%%%%%%%%%%%%

By Proposition \ref{pr:synchronous-relations}  every synchronous relation is Presburger definable.
Consequently,  
 given a binary synchronous relation $R\subseteq \N\times \N$ it is decidable whether or not it is an order (resp. linear order),
whether or not for a given integer $n$ it has a chain resp. an antichain of size less than or equal to $n$, whether or not 
it is $N$-free etc\ldots. Here we tackle problems that do not seem to be expressible in  Presburger theory.

 \begin{proposition}
 It is decidable whether or not a synchronous  order has an infinite  chain, resp. an infinite antichain.
 
 If there is no infinite antichain, the lengths of the antichains are bounded by 
  $2n+2$ where $n$ is the number of states of the subautomaton $\+B$.
 
 \end{proposition}

  \begin{proof}  We may assume that the relation is complete. Indeed, let $E$ be the complement of the support of $R$.
Then the synchronous relation $R\cup \text{supp}(R)\times E$ is complete 
 and it has infinite chains if and only if so does $R$.

We use the notations introduced in paragraph \ref{ss:binary-case}
  and in particular we let $t$ be the transient of the subautomaton $\+B$.
Since a descending chain is ascending for the inverse ordering and since the inverse of a synchronous order on $\N$ is also synchronous  it suffices to consider ascending chains.
%.
We claim that there exists an infinite chain  if and only if some $\aad_{i}$ for  $t\leq  i<n$ 
recognizes $p$ (the common period),
i. e., $\theta_{i}(q_{i},p)$ is final.
It is clearly sufficient since in this case consider some integer $k$ taking $q_{0}$ to $q_{i}$ in $\+B$. 
Then $q_{0}\cdot (k+p)= q_{i}$ and thus $k\prec k+p\prec k+2p\cdots $.
Conversely, if there exists an infinite chain there exists an infinite  chain that is increasing  for the natural order on $\N$.
Then there exist $t\leq k<\ell$ such that $\ell-k$ is a multiple of $p$, say $\alpha p$. Let $q_{0}\cdot k=q_{i}$. Then  $k\prec \ell$ implies $\theta_{i}(q_{i}, \alpha p)$ is a final state of 
$\aad_{i}$ thus $\theta_{i}(q_{i},  p)$ is a final.

\medskip 

We now turn to antichains.
By lemma \ref{le:finite-complement}    and corollary \ref{cor:infinite-complement}  
we may assume that the relation is  complete.
We claim that 
there exists an infinite antichain if and only if there exits an integer $i$ such that $\lambda_{i}(q_{i}, p)$ and 
$\theta_{i}(q_{i}, p)$ are non final in $\aag_{i}$ and $\aad_{i}$ respectively. The condition is sufficient. 
Indeed,  because $R$ is complete  any two elements  elements $k<\ell$   taking $q_{0}$ to $q_{i}$ 
 differ by a multiple of $p$. Then $(k,\ell)\not\in R$ and $(\ell,k)\not\in R$
 implying that  $p$ is not a final state of $ \aag_{k}$  and $\aad_{k}$.
Conversely, if there exist infinitely many pairwise incomparable elements there exist infinitely any elements greater than $t$ 
taking $q_{0}$ to the some $q_{i}$. For two such elements $k$ and $\ell$, we have  $(k,\ell)\not\in R$ and $(\ell,k)\not\in R$,   $\lambda_{i}(q_{i}, p)$ and thus
$\theta_{i}(q_{i}, p)$ are nonfinal in $\aag_{i}$ and $\aad_{i}$ respectively. 

\medskip Now, suppose there is no infinite antichain. We claim that an antichain  has less than $2n+2$ elements. Indeed, in an antichain containing  $2n+2$ elements  there are two elements $k<\ell$ with  $q_{0}\cdot k= q_{0}\cdot \ell$ is not a transient state and thus $\ell-k$ is a multiple of $p$. Then 
$\lambda_{i}(q_{i}, p)\not\in \aag_{i}$ and $\theta_{i}(q_{i}, p)\not\in \aad_{i}$ but then the previous discussion shows that 
there exists an antichain of infinite length.

\end{proof}

 Observe that infinite antichains exist, e.g., $\{(2n,2n+1)\mid n\in \N\}$ so that the question on the upper bound on antichains makes sense. Also there is a departure with orders on $\N\times\N$: the product of the usual order on $\N$ has no infinite antichain but has antichains of arbitrary lengths.
 
 \begin{proposition}
 Given two synchronous automata defining linear orders, it is decidable whether or not these orders are equivalent.
\end{proposition}
 
  \begin{proof} We assume first that the orders are complete.
By Proposition \ref{pr:reduced-expression} it suffices to prove that we can effectively associate with a synchronous 
automaton defining a linear order on $\N$ an expression of the form \ref{eq:poor-sum}. Using the notation of Theorem  \ref{th:char-linear}, the set 
$\N$  is the finite and disjoint union of the singletons $\{0, \ldots, t-1\}$ and ascending and descending chains \ref{ascending} and \ref{descending}. By proceeding as in Proposition \ref{pr:charac-poor-linear-orders} it suffices to relate them pairwise.

\medskip Consider a chain $C_{\alpha}$ as in \ref{ascending} or \ref{descending} (with $t\leq \alpha <n$) and an arbitrary $\gamma\in \N$. Let $k$ be the least integer such that 
$\alpha+ kp>\gamma$. Then
\begin{equation}
\label{eq:comparison}
\begin{array}{l}
\alpha+ (k+1) p\succ \gamma \Leftrightarrow \alpha+ \ell  p\succ \gamma \text{ for all } \ell\geq k+1\\
\alpha+ (k+1) p\prec \gamma  \Leftrightarrow \alpha+ \ell  p\prec \gamma \text{ for all } \ell\geq k+1
\end{array}
\end{equation}

This implies that it is possible to determine the integer $r$ such that $\gamma$ lies between $\alpha + rp$ and $\alpha + (r+1)p$
and solves the problem of how a singleton and a chain relate.

\medskip We turn on to the problem of determining how  two chains  $C_{\alpha}: \alpha,   \alpha +p ,    \cdots $ and 
$C_{\beta}: \beta,   \beta +p,    \cdots$ relate. We assume $t\leq \alpha<\beta< n$.  
Because it is possible to determine how an element relates to a chain, it is sufficient to compare 
two tails of $C_{\alpha}$ and $C_{\beta}$ where at most finitely many  only  first elements are missing. We use the equivalences \ref{eq:a2b}, \ref{eq:ab} and \ref{eq:deflating}.

\nl Case 1. Two ascending chains 

$$
C_{\alpha}: \alpha\prec   \alpha +p \prec   \cdots \text{ and }  C_{\beta}: \beta\prec   \beta +p \prec   \cdots 
$$
\nl \fbox{Case 1.1.}  $\beta  \succ \alpha  + 2 p  $  implies 
$C_{\beta}\succ C_{\alpha}\setminus \{\alpha, \alpha  + p\}$ by \ref{eq:a2b}

\nl \fbox{Case 1.2.}  $\alpha   \prec \beta  +p $ and $\beta \prec \alpha +2p$ implies  because of \ref{eq:a2b} and \ref{eq:ab}.
$$\alpha   \prec \beta  +p \prec \alpha + 3 p \prec \beta + 4p \prec \alpha + 6 p\prec \cdots
$$ which is an interleaving of $C_{\alpha}$
and $C_{\beta}$.

\nl \fbox{Case 1.3.}  $\alpha   \succ \beta  +p $  implies 
 $C_{\alpha} \succ C_{\beta} \setminus \{\beta\}$ by \ref{eq:ab}

\nl Case 2. Two descending chains 
$$
C_{\alpha}: \alpha\succ   \alpha +p \succ   \cdots \text{ and }  C_{\beta}: \beta\succ   \beta +p \succ   \cdots 
$$
\nl \fbox{Case 2. 1.}  $\alpha  \prec \beta + p$ implies 
$C_{\alpha}\prec C_{\beta}\setminus \{\beta\}$ by \ref{eq:ab}.

\nl \fbox{Case 2. 2.}  $\alpha \succ  \beta + p  $ and $\beta\succ \alpha +2p$   implies
$\beta + p \succ \alpha +3p\succ \beta +4 p \succ \alpha +6p \succ \cdots  $ which is an interleaving of $C_{\alpha}$
and $C_{\beta}$ because of \ref{eq:a2b} and \ref{eq:ab}.

\nl \fbox{Case 2. 3.} $\beta\prec \alpha +2p$ implies $\beta\prec C_{\alpha}$ thus $C_{\beta}\prec C_{\alpha}\setminus \{\alpha, \alpha +p\}$  by \ref{eq:ab}.

\nl Case 3. An ascending  and a descending chain
$$
C_{\alpha}: \alpha\prec   \alpha +p \prec   \cdots \text{ and }  C_{\beta}: \beta\succ   \beta +p \succ   \cdots 
$$
%
%\break 

\nl \fbox{Case 3.1. } $\beta\prec \alpha +2p $ 
which yields 
$C_{\beta} \prec C_{\alpha}\setminus \{\alpha, \alpha +p\}$ because of \ref{eq:a2b}

\nl \fbox{Case 3.2.}  $\alpha\prec  \beta +p$ and  $\beta\succ \alpha +2p $ implies $\alpha \prec C_{\beta}\setminus \{\beta\} $
and $\beta\succ C_{\alpha}\setminus \{\alpha, \alpha +p\} $. Assume 
 $\beta +kp \prec \alpha +\ell p$. Since $C_{\beta}$ is decreasing we may assume $k-\ell\geq 1$, thus 
 $\beta + (k-\ell) p \prec \alpha$  because of \ref{eq:deflating} which contradicts $\alpha \prec C_{\beta}$.
 Thus $C_{\alpha}  \prec C_{\beta}$.

\nl \fbox{Case 3.3.}  $\beta +p \prec \alpha  $
 implies 
$C_{\beta}\setminus \{\beta\} \prec C_{\alpha}$ because of \ref{eq:ab}.

%\break 

\nl Case 4. A descending   and an ascending  chain
$$
C_{\alpha}: \alpha\succ    \alpha +p \succ   \cdots \text{ and }  C_{\beta}: \beta\prec    \beta +p \prec   \cdots 
$$

\nl \fbox{Case 4.1. } $\beta\succ \alpha + 2p$  
then 
$C_{\beta} \succ C_{\alpha} \setminus \{\alpha, \alpha +p\}$ because of \ref{eq:a2b}

\nl \fbox{Case 4.2.}  $\alpha\prec  \beta + p$ implies  $C_{\alpha} \prec C_{\beta}\setminus \{\beta\}$ because of \ref{eq:ab}

\nl \fbox{Case 4.3.}  $\alpha\succ  \beta + p$ and $\beta\prec \alpha + 2p$. Thus $\alpha \succ C_{\beta}\setminus \{\beta\}$
and $\beta \prec C_{\alpha}\setminus \{\alpha, \alpha +p\}$. Assume $\alpha +kp\prec \beta +\ell p$ with $k<\ell$ without loss of generality. 
Then $\alpha \prec \beta + (\ell-k)p$ contradicting $\alpha \succ C_{\beta}$
thus $C_{\beta} \prec C_{\alpha}$

\bigskip Now we consider the general case where $\prec_{1}$ and $\prec_{2}$ are two nonncessarily complete orders
of type $\tau_{1}$ and $\tau_{2}$. Set  $E_{1}= \text{supp}(\prec_{1})$, $E_{2}= \text{supp}(\prec_{2})$.  If the two orders have the same type, then either 
they  have both a maximal element or they have both no maximal element. This property can be verified because 
an order has a maximal element if and only if the complement of the following set is nonempty.
$$
\{k+\ell\mid \ell\in \aag_{k}\} \cup \{k\mid \exists \ell\in \aad_{k}\} 
$$
By  lemma \ref{le:finite-complement}  we may assume that  $E_{1}$ and $E_{2}$ are infinite. 
If the two orders have a maximal element  then using lemma \ref{le:linear-on-regular}
we can complete the two orders and obtain  orders of  types $\tau_{1}+ \omega^*$ and $\tau_{2}+ \omega^*$.
Then $\tau_{1}+ \omega^*=\tau_{2}+ \omega^*$. if and only if $\tau_{1}=\tau_{2}$. Similarly if the 
 two orders have no maximal element we can complete the two orders and obtain  
 orders of order types $\tau_{1}+ \omega$ and $\tau_{2}+ \omega$.
Then $\tau_{1}+ \omega=\tau_{2}+ \omega$ if and only if $\tau_{1}=\tau_{2}$.  

 \end{proof}

\end{document}